# Giant enhancement of spin detection sensitivity in (Ga,Mn)As/GaAs Esaki diodes

Junichi Shiogai[1], Mariusz Ciorga[2], Martin Utz[2], Dieter Schuh[2], Makoto Kohda[1], Dominique Bougeard[2], Tsutomu Nojima[3], Junsaku Nitta[1] and Dieter Weiss[2]

[1]*Department of Materials Science, Tohoku University, 980-8579 Sendai, Miyagi, Japan*
[2]*Institute of Experimental and Applied Physics, University of Regensburg, D-93040 Regensburg, Germany*
[3]*Institute of Materials Research, Tohoku University, 980-8577 Sendai, Miyagi, Japan*

We investigate the correlation between spin signals measured in three-terminal (3T) geometry by the Hanle effect and the spin accumulation generated in a semiconductor channel in a lateral (Ga,Mn)As/GaAs Esaki diode device. We systematically compare measurements using a 3T configuration, probing spin accumulation directly beneath the injecting contact, with results from nonlocal measurements, where solely spin accumulation in the GaAs channel is probed. We find that the spin signal detected in the 3T configuration is dominated by a bias-dependent spin detection sensitivity, which in turn is strongly correlated with charge-transport properties of the junction. This results in a particularly strong enhancement of the detected spin signal in a region of increased differential resistance. We find additionally that two-step tunneling via localized states (LS) in the gap of (Ga,Mn)As does not compromise spin injection into the semiconductor conduction band.

All-electrical generation, manipulation and detection of spin-polarized electrons in semiconductors are key prerequisites for the realization of spin-based electronic devices.[1,2] In recent years there has been considerable progress in understanding the basic processes governing electrical spin injection from a ferromagnet (FM) into a semiconductor (SC), with numerous theoretical and experimental contributions.[3–10] Despite this, the large spin signals[11], measured particularly in Si- and Ge-based devices[12–14], withstand so far straightforward explanation and go well beyond (i.e., orders of magnitude) the commonly accepted standard model of spin injection.[3,4] These large values have been found using a three-terminal (3T) method of spin detection, with one single magnetic contact used to inject and detect spin accumulation. An initially proposed model explained the observed giant spin signal enhancement in terms of an enhanced spin accumulation generated solely in states localized at the FM/SC interface[11]. The enhancement is then driven by the large resistance between localized states (LS) and the SC channel, due to a depletion zone in the interface region. Such a scenario would however impede actual spin injection into the SC channel itself[11,15], questioning also the applicability of the 3T method to detect spin accumulation in the SC channel. This constraint was relaxed in an extension of the LS model, allowing for direct tunneling of electrons between FM and SC conduction band, suggesting that direct band-to-band tunneling and double-step tunneling, involving LS, occur simultaneously.[16] On the other hand, it was shown in some experiments that the spin signal can be enhanced even in the absence of a depletion region.[14,17]

What has been missing so far is a systematic comparison between 3T and NL measurements on systems showing large signals in 3T configuration, needed in order to unambiguously establish the correlation between the 3T signal and the actual spin accumulation in the channel. What also has been overlooked in the recent discussion is the detection sensitivity of spin detecting contacts. It was shown theoretically by Chantis & Smith[18], and observed experimentally by Crooker *et al.*[19] that a current-biased spin detector has its sensitivity dramatically changed compared to a non-biased case. This makes charge transport through the detecting contact, in particular in the presence of any non-linearity, a very important factor. Because one uses a biased contact as spin detector in 3T configuration, these effects should be taken into account for analyzing the measured signals.

In this paper we employ a lateral (Ga,Mn)As/GaAs spin injection device as a test bed [7,20-24] to investigate the effects describe above. The use of Esaki diodes as spin sensitive contacts gives us the unique opportunity to tune the relative contribution of direct and two-step tunneling via localized states in the gap of (Ga,Mn)As in a single device by simply changing the bias across the junction.[25] We show that (i) tunneling through localized states, while generating a spin accumulation there, does not affect spin accumulation in the conduction band of GaAs and (ii) that the detection sensitivity is strongly affected by the non-linearity of the current-voltage (*I-V*) characteristic of the contact.

A schematic of a typical spin injection device is depicted in Fig. 1(a). The device is patterned into a 50-μm-wide, [110] oriented mesa by standard photolithography and wet chemical etching techniques. The corresponding wafer consists of a semi-insulating (001) GaAs substrate, a 300 nm GaAs buffer layer, a 500 nm AlGaAs/GaAs superlattice, 0.8 μm *n*-GaAs, 0.2 μm $n^+$-GaAs, a 15 nm $n^+ \rightarrow n^{++}$-GaAs transition layer, 8.0 nm $n^{++}$-GaAs, a 2.2 nm AlGaAs diffusion barrier, and 50 nm (Ga,Mn)As. The doping concentration of the GaAs layers are $n = 2\times10^{16}$ cm$^{-3}$, $n^+ = 6\times10^{16}$ cm$^{-3}$, $n^{++} = 6\times10^{18}$ cm$^{-3}$, respectively. Ferromagnetic contacts, aligned along [1$\bar{1}$0], were defined by electron beam lithography, Au/Ti evaporation and lift-off. Contact 2 is 4 μm and the other contacts (3 – 6) are 0.5 μm wide. The center-to-center spacing between neighboring contacts is 5 μm, and *L* is the distance between injector and detector. Large reference contacts (150 μm x 150 μm) were defined at the end of the mesa by photolithography and Au/Ti evaporation. Finally, the (Ga,Mn)As layer and the highly doped GaAs layer were removed between the contacts by reactive ion etching so that the current flow is confined to the low-doped GaAs channel. Contact 2 was usually used as injector, the others as detectors. A non-equilibrium spin accumulation generated underneath the injector by driving a current $I_{\text{inj}}$ between ferromagnetic injector and reference contact 1 can then be probed either nonlocally or using the 3T method.

Let us first discuss the *I-V* characteristic of contact 2, shown in Fig. 1(b). The curve was taken in 3T configuration, i.e., $I_{\text{inj}}$ flows across the (Ga,Mn)As/GaAs Esaki diode and the voltage drop across the junction, $V_{3T}$, is measured between contact 2 and reference contact 7. The current through an Esaki diode consists of different contributions from: (i) direct tunneling between the valence band of *p*-(Ga,Mn)As and the conduction band of *n*-GaAs; (ii) tunneling through localized states in the bandgap (constituting the so-called excess current[25]) and (iii) thermal transport across the built-in potential. The component (iii), not interesting for spin injection, is dominating at high forward bias. At reverse bias and for small forward bias the component (i) dominates the current as electrons tunnel from (Ga,Mn)As into GaAs (reverse bias) or in the opposite direction (forward bias). The latter case is schematically shown in the upper inset of Fig. 1(b). A further increase of the forward bias removes the overlap of the bands, suppressing the component (i). For an ideal Esaki diode this would lead to a vanishing current (see the dashed curve in the Fig. 1(b)). In real devices, however, the component (ii) dominates in this regime and is responsible for a non-zero tunnel current.[25] In our devices the importance of this process is manifested by a very shallow Esaki dip at about 0.4 V in Fig. 1(b). This type of Esaki dip is observed in all our (Ga,Mn)As/GaAs junctions and indicates the presence of a large excess current. This is not surprising as (Ga,Mn)As, grown at low temperatures, contains a high density of localized states in the gap[26–28] which supports two-step (or multi-step) tunneling. This situation is depicted in the lower inset of Fig. 1(b), showing electrons tunneling from the conduction band either into localized states or directly into the valence band. For our further discussion it is important to note that the *I-V* characteristic of the Esaki diode, while nearly linear in the direct tunneling regime (i) becomes highly non-linear in regime (ii). Thus, by tuning the bias voltage between the red and the blue point in Fig. 1(b), both, the ratio of direct and indirect tunneling currents and the degree of non-linearity of the *I-V* characteristics are widely changed. The junction resistance $R_{3T}A$, with the contact area $A$, varies from 0.5 to 1.5 MΩμm², depending on the injected current. In the whole range it exceeds the lower limit for efficient spin injection[3], given by $R_{3T}A \gg r_s^{ch} = \rho_N \lambda_{sf} \approx 4.0$ kΩμm². Here, $\rho_N$ and $\lambda_{sf}$ are the resistivity and the spin diffusion length of the GaAs channels, respectively, and $r_s^{ch}$ is the effective spin resistance of the channel.

In our NL measurements the four 0.5 µm wide contacts 3 – 6 are used as nonlocal spin detectors probing pure spin currents flowing from the injector towards the detectors. According to the standard drift-diffusion model, the spin accumulation at the injection point, $\mu_s(0) = -P_{inj}jr_s^{ch} = -P_{inj}j\rho_N\lambda_{sf}$, gives rise to the following NL voltage at the detection point $x = L$ [4–7,29]

$$V^s(L) = -P_{det}\mu_s^{ch}(L)/2 = \pm(P_{inj}P_{det}I_{inj}\lambda_{sf}\rho_N/2S)\exp(-L/\lambda_{sf}), \quad (1)$$

where $I_{inj}$ is the spin injection current, $S$ is the cross-sectional area of the nonmagnetic channel and $P_{inj(det)}$ is the tunneling spin polarization (TSP) of the injector (detector). The + (−) sign stands for the parallel (antiparallel) magnetization alignment of injector and detector. The magnetization configuration is switched from parallel to anti-parallel (or vice versa) by sweeping an in-plane magnetic field $B_x$ oriented parallel to the contacts. The switching results in a voltage jump $\Delta V_{nl} = 2 \cdot V^s(L)$ which is a direct measure of the generated spin accumulation. Alternatively, the Hanle effect is used, where an out-of-plane magnetic field $B_z$ causes a precession of the in-plane electron spins, which results in a decay of the Hanle amplitude $\Delta V_{nl}^{Hanle} = V^s(L)$ as a function of $B_z$.

Spin detection in 3T configuration relies on the fact that spins accumulated at the interface between ferromagnet and semiconductor increase the voltage drop across the junction.[11,16] The spin-related contribution to $V_{3T}$ is typically described by Eq. (1) assuming $L = 0$, $P_{inj} = P_{det}$, and can be determined by Hanle measurements. The suppression of spin accumulation at finite $B_z$ results in a reduction of the measured $V_{3T}$ voltage with the signal amplitude $\Delta V_{3T}^{Hanle} = V^s(L = 0)$.

Spin signals measured in the regime of direct tunneling and indirect tunneling (two-step or multi-step) are shown in Fig 1(c) and 1(d), respectively. Clear nonlocal spin valve and Hanle signals, observed for both bias regimes, indicate that a spin current flows along the channel, i.e., spin accumulation in the conduction band of GaAs occurs. From distance-dependent measurements (not shown) we determine the spin diffusion length $\lambda_{sf} = 6.0$ μm; using Eq. (1) we extract $P_{inj} = 0.641$ at $V_{3T} = 0.043$ V and $P_{inj} = 0.194$ at $V_{3T} = 0.336$ V in good agreement with our previous work.[7] Comparison of the nonlocal spin resistances $\Delta R_{nl}^{Hanle} = \Delta V_{nl}^{Hanle}/I_{inj}$ for $L = 5$ μm (top traces in Fig. 1(c) and 1(d)) shows that this Hanle signal drops from 9 Ω in the direct tunneling regime to 3 Ω for tunneling via localized states. In contrast, the 3T-signal (bottom traces in Fig. 1(c), 1(d)) increases from $R_{3T}^{Hanle} = 21 \Omega$ in the low bias regime to $157 \Omega$ in the impurity-assisted tunneling regime. Using Eq. (1) and the parameters extracted from nonlocal measurements we can estimate the expected Hanle signal: For the direct tunneling regime ($V_{3T} = 0.043V$) we calculate $\Delta R_{3T}^{Hanle} = 20\Omega$ while we obtain $\Delta R_{3T}^{Hanle} \approx 1.8\Omega$ for the indirect tunneling ($V_{3T} = 0.336V$). While calculated and measured Hanle signals are nearly the same in the direct tunneling regime, where the *I-V* characteristic is nearly linear, they differ by about two orders of magnitude in the indirect tunneling regime, i.e., when the *I-V* characteristic becomes non-linear.

To investigate this discrepancy in more detail we systematically studied the dependence of the spin signals on the bias voltage/current across the injector. The results are summarized in Fig. 2 where we plot both $\Delta R_{nl}^{Hanle}$ and $\Delta R_{3T}^{Hanle}$ as a function of *V₃T*. The nonlocal resistance $\Delta R_{nl}^{Hanle}$ decreases monotonically with increasing bias in both directions, ascribed to a decrease of *P_{inj}*.[7] The behavior of the 3T-Hanle signal $\Delta R_{3T}^{Hanle}$ is strikingly different from the theoretical prediction $\Delta R_{3T}^{th} = P_{inj}^2 \lambda_{sf} \rho_N / 2S$, which is plotted as a dashed curve in Fig. 2. This prediction is based on Eq. (1) using the parameters extracted from nonlocal measurements and thus follows qualitatively $\Delta R_{nl}^{Hanle}$. $\Delta R_{3T}^{Hanle}$, in contrast, slowly increases for positive bias, reaches a plateau and then rises again to reach a maximum at the Esaki dip position. A further increase of the voltage rapidly decreases the signal. For reverse bias the signal rapidly drops to zero before changing its sign at $V_{3T} = -0.1$ V. The behavior at low positive and low negative bias resembles well nonlocal experiments on the Fe/GaAs system with a biased detector, interpreted in terms of bias dependence of the detector's sensitivity.[18,19] The sensitivity is defined as a change in a voltage drop $\Delta V$ across the biased FM/SC interface as a result of spin accumulation $\Delta \mu_s$ generated in the SC. Its bias dependence can be quite different from bias dependence of $P_{\text{inj(det)}}$ and stems from the dependence of the density of spin-polarized carriers underneath the detector on the electric field in the channel and at the interface. As a result the spin signal is expected to be enhanced for spin extraction ($I > 0$) and suppressed for spin injection ($I < 0$) case. This is exactly what we observe in the experiments as $\Delta R_{3T}^{Hanle} > \Delta R_{3T}^{th}$ for the former and $\Delta R_{3T}^{Hanle} < \Delta R_{3T}^{th}$ for the latter, see Fig. 2.

Let us now discuss the huge enhancement of the local signal in the vicinity of the Esaki dip, displayed also in the inset of Fig. 2, i.e., in a regime where the tunneling current is dominated by the excess current. Because we do not observe either enhancement or suppression in the NL signal, one can conclude that the spin current flowing in the channel (i.e., in the conduction band of GaAs) and thus the spin accumulation which builds up underneath the nonlocal detector is not affected by the excess current. There are then two possible mechanisms which can account for the enhancement of the 3T signal. The first one involves spin injection into localized states with a higher spin effective resistance $r_s^{ls}$ than the one in the channel $r_s^{ch}$.[11] This would result in a higher spin accumulation underneath the injecting contact and would thus dominate the measured $\Delta R_{3T}^{Hanle}$ without changing the spin current in the channel. A second possible mechanism is based on an increased sensitivity of spin detection in the highly non-linear region of the Esaki dip. It can be explained as follows. Consider the voltage drop across the junction $V_{3T}$ in the presence of the constant injection current $I$. It contains the contribution $V^s = -(P_{det}/2)\mu_s$ stemming from the generated spin accumulation $\mu_s$ and for $I > 0$ it can be written as $V_{3T}(\mu) = IR(V_{3T}) + (P_{det}/2)|\mu_s|$, taking into account that the interface resistance $R_{3T} = R(V_{3T})$ is also voltage dependent. In the Hanle experiments the spin accumulation is reduced by $\Delta\mu_s$ due to the applied out-of-plane field $B_z$, resulting in spin precession and dephasing. A condition of the constant current requires readjustment of the voltage across the junction by $\Delta V_{3T}$ if the spin accumulation changes by $\Delta\mu_s$. This adjustment, which constitutes the detection sensitivity addressed above, is readily obtained by taking a derivative of the above expression with respect to $\mu_s$ and results in $\Delta V_{3T} = (1 - IdR_{3T}/dV_{3T})^{-1}P_{det}\Delta\mu_s/2$ or, equivalently, $\Delta V_{3T} = (dV_{3T}/dI)/(V_{3T}/I)\Delta\mu_s$, in agreement with the expression derived in Ref. 18. This means that the spin detection sensitivity is amplified by the ratio of the differential resistance and the interface resistance, a measure of non-linearity of the *I-V* curve. The 3T Hanle signal $\Delta V_{3T}^{Hanle}$, measured as the voltage change $\Delta V_{3T}$ due to full depolarization of the spin accumulation, is then expected to be proportional to that ratio, This proportionality can be seen in Fig. 2. by comparing the $\Delta R_{3T}^{Hanle}$ signal with $(dV_{3T}/dI)/(V_{3T}/I)$ (top panel), calculated from the *I-V* curve (see Fig. 1). The 3T signal is clearly enhanced in the region of high non-linearity.

To disentangle these two contributions we performed spin-valve measurements with the setup shown schematically in Fig. 3(a), allowing to directly measure the spin detection sensitivity.[19] Now, contact 2 serves as a biased nonlocal detector of the spin accumulation generated in the GaAs channel by applying a small ac current bias with frequency 17 Hz to contact 3. The nonlocal voltage $V_{nl}^{ac}$ is then measured as a function of a dc current bias applied to contact 2, used to tune the voltage drop $V_{3T}^{dc}$. In Fig. 3(c) we show the nonlocal spin-valve signal $V_{nl}^{ac}$, obtained by sweeping $B_x$, at different values of $V_{3T}^{dc}$ marked in Fig. 3(b). The spin-valve amplitude $\Delta V_{nl}^{ac}$, which is now a direct measure of the spin detection sensitivity, strongly depends on the applied $I_{dc}$ in a similar manner as the 3T signal: a suppression and sign reversal is observed at negative bias while a strong amplification is observed at the Esaki dip.

In Fig. 3(d) we compare the bias-dependent enhancement of the spin signals observed in both configurations, i.e., $\Delta R_{3T}^{Hanle}$ in Fig. 2 and $\Delta V_{nl}^{ac}$ in Fig 3(c). The comparison is done by introducing an enhancement factor. In case of the biased nonlocal detector the enhancement factor is calculated as $\Delta V_{nl}^{ac}/\Delta V_{nl}^{ac}(I_{dc}=0) \times P_{det}(I_{dc}=0)/P_{det}$, and plotted as red data points in Fig. 3(d). Here we take into account that $P_{det}$ decreases with increasing bias current $I_{dc}$. In 3T case the enhancement factor is defined as $\Delta R_{3T}^{Hanle}/\Delta R_{3T}^{th}$, i.e., the ratio of the blue and the green dashed traces in Fig. 2. In the regime of direct tunneling (i), plotted in the inset of Fig. 3(d), signals in both configurations show good qualitative and quantitative agreement, i.e., enhancement in the forward bias, reaching a factor of 4 at 0.2 V, and suppression and sign reversal in the reverse bias. This behavior is fully consistent with the results of Refs. 1 19. As a result of non-linearity[18] in the Esaki dip region the detection sensitivity is further enhanced, reaching a factor of about 40. The enhancement of the 3T signalin this region is, however, still two times higher, suggesting that the excess current generates also spin accumulation in gap states that contributes to the signal. We conclude therefore, that the enhanced 3T signal, although having the contribution from LS, originates predominantly from the increased sensitivity to detect a conduction band spin accumulation. This enhancement is strongly correlated with charge-transport through the interface, namely, the non-linearity of the *I-V* characteristic of the junction.

In summary, we studied the correlation between 3T spin signal and spin accumulation in the semiconductor channel probed in non-local geometry. Our first fully comparative 3T- and NL-Hanle experiments show that tunneling through LS does not affect spin injection into the conduction band of a SC channel, and that the 3T method can be used to detect spin accumulation in the channel. One has to be very careful, however, while extracting the actual magnitude of the generated spin accumulation, as the measured signal is dominated by the bias-dependent sensitivity of spin detection.[18,19] As a result, Eq. (1) cannot longer be used to describe the measured spin signal when the detector is biased, like in the case of the 3T method. This aspect of the 3T spin detection was hitherto not taken into account although some experiments on Si devices show correlation between measured spin signals and tunnel resistance[30,31] or differential resistance[32] of the junction. Although our experiments were conducted on spin Esaki diode devices, we find the results are quite general. Especially the possibility to amplify the tiny nonlocal spin signals by engineering a tunnel barrier in the detector in a way that it shows a high $(dV_{3T}/dI)/(V_{3T}/I)$ ratio in the detector can be of a significant importance for the development of future spintronic devices.

We thank H. Jaffrès and J. Fabian for very fruitful discussions. This work was partly supported by the German Science Foundation (DFG) via SFB 689, the Japan-Germany Strategic International Cooperative Program (Joint Research Type) from JST and DFG (FOR 1483) and Grants-in-Aid from JSPS 22226001 and 24684019.

**Figure captions**

**Fig. 1** (Color online) a) Multi-terminal spin injection device for three-terminal and nonlocal detection. b) Current-voltage (*I-V*) characteristic of spin Esaki contact 2. The dashed line shows schematically the *I-V* characteristic of an ideal Esaki diode in the absence of excess current. Insets show schematically direct (upper) and indirect (lower) tunneling processes. c) Nonlocal (upper panel) and three-terminal (lower panel) voltages in the regime of direct tunneling ($I_{inj}$=5 µA). Colored traces are plotted as a function of out-of-plane field $B_z$ (Hanle signal $\Delta R^{Hanle}_{37(3T)} = \Delta V^{Hanle}_{37(3T)}/I_{inj}$) while the thin grey lines are up- and down-sweeps of the in-plane field $B_x$. d) As c), but for $I_{inj}$=60 µA, i.e., in the excess current regime.

**Fig. 2** (Color online) Bottom panel: Spin accumulation signal for 3T-Hanle signal $\Delta R^{Hanle}_{3T} = \Delta V^{Hanle}_{3T}/I_{inj}$, detected at contact 2 and 4T-Hanle signal $\Delta R^{Hanle}_{nl} = \Delta V^{Hanle}_{nl}/I_{inj}$, measured at contact 3 and plotted as a function of bias voltage $V_{3T}$. Open circles of the 3T-Hanle signal indicate negative sign in $\Delta R^{Hanle}_{3T}$. The dotted line shows the expected value for $\Delta R^{th}_{3T}$ calculated from Eq. (1) for $L = 0$ using parameters extracted from nonlocal measurements. The inset shows the enhancement factor defined

as $\Delta R_{3T}^{Hanle}/\Delta R_{3T}^{th}$. Top panel: the ratio of differential resistance $dV_{3T}/dI$ and junction resistance $R_{3T} = V_{3T}/I$ calculated from the *I-V* characteristic shown in Fig 1.

**Fig. 3** (Color online) a) Schematic of the circuit used for ac+dc measurements to extract the bias-dependent detection sensitivity; b) Corresponding *I-V* curve of contact 2; c) spin-valve signal traces measured at nonlocal detector 2, with spin accumulation generated by contact 3 using an excitation current $I_{ac}$ = 4.7 µA , and dc bias voltages $V_{3T}^{dc}$ as marked in b). Curves are shifted for clarity. d) Enhancement (suppression) of the NL and 3T signals by applied dc bias. Inset: data from low bias-measurements, i.e., in the regime of direct tunneling. For details see text.

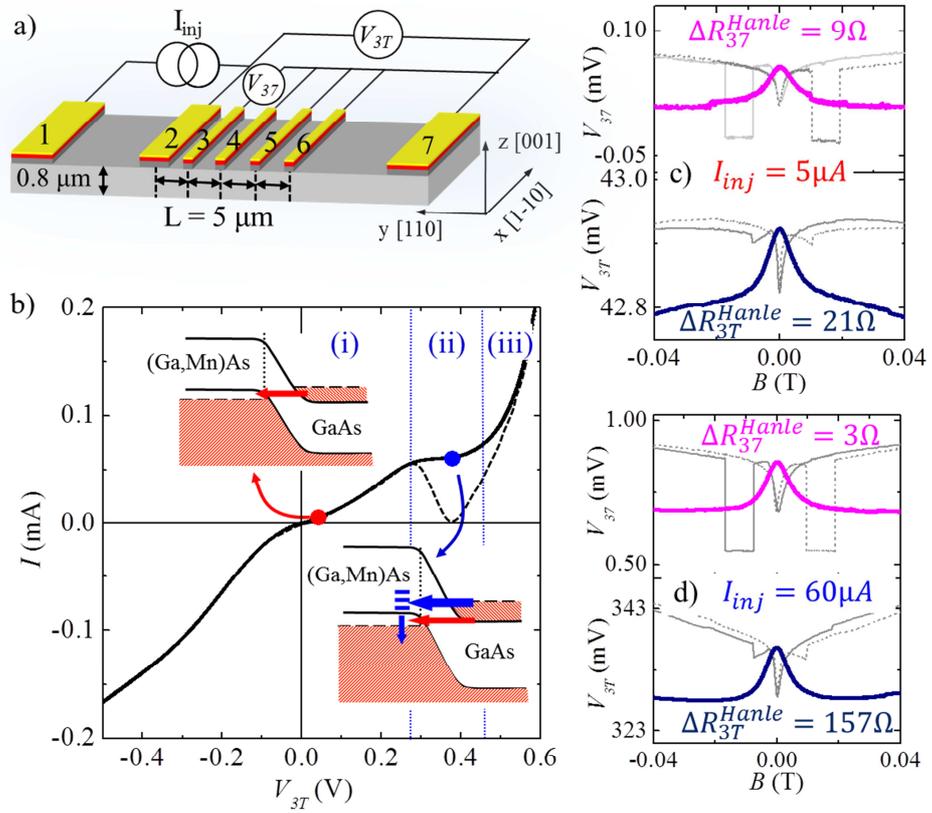

Fig. 1

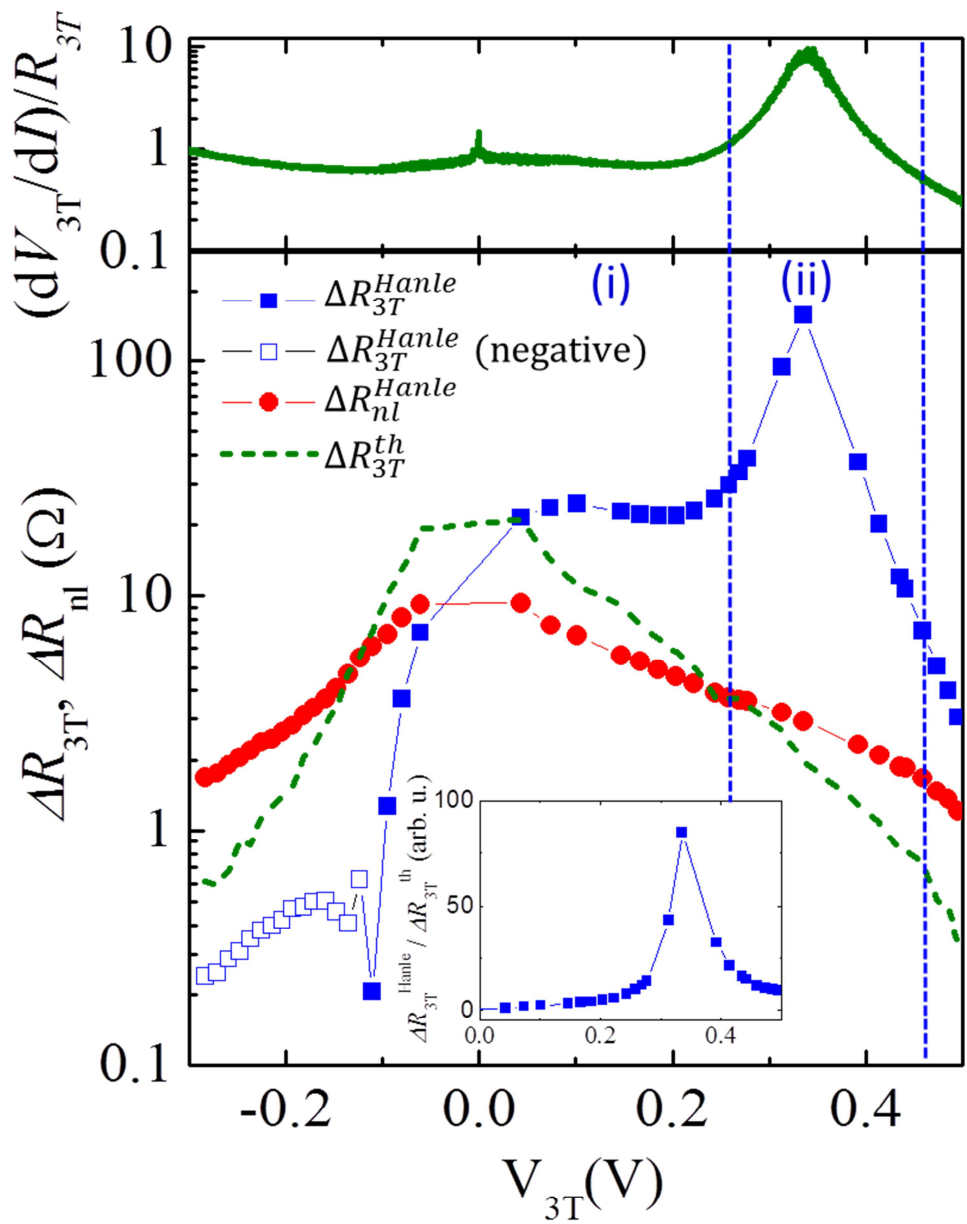

Fig. 2

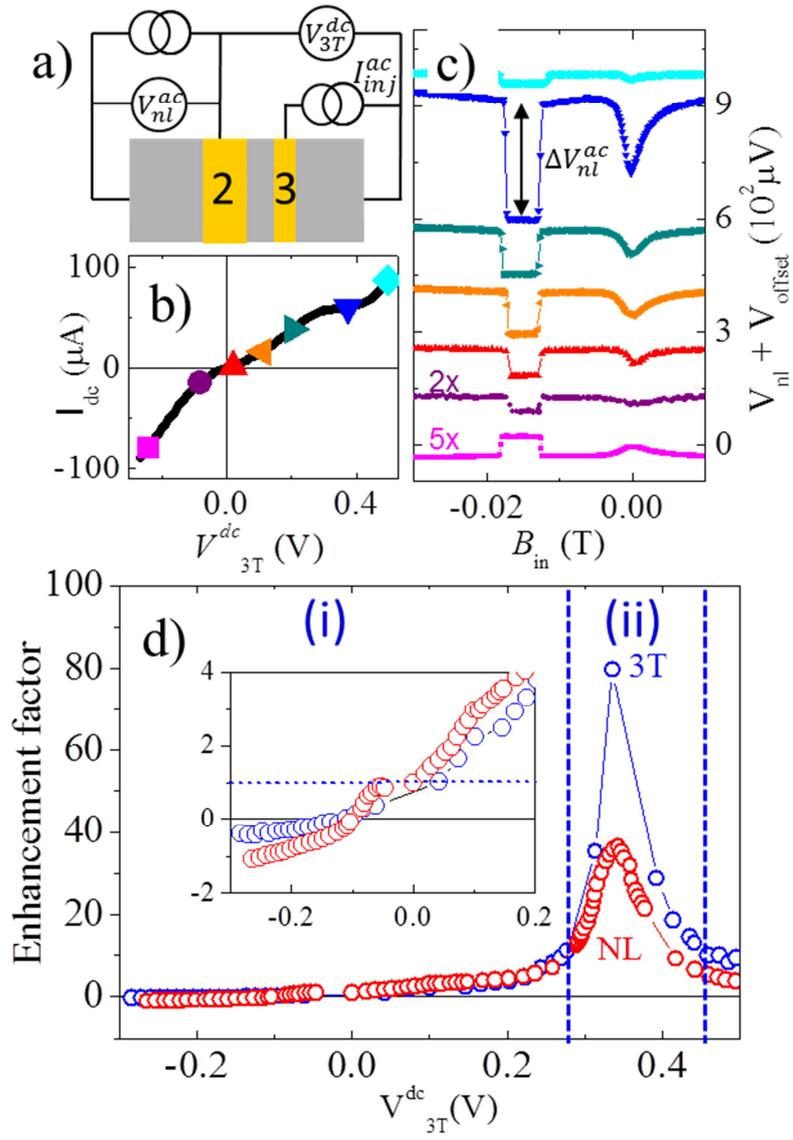

Fig. 3.